\documentclass[aps,prb,twocolumn,showpacs,groupedaddress]{revtex4-1}
\usepackage{amssymb}
\usepackage{hyperref}
\usepackage{color}
\usepackage{graphicx}
\usepackage[caption=false]{subfig} 
\usepackage{amsmath}

\begin{document}

\title[]{Extremely large non-saturating magnetoresistance and ultrahigh mobility due to topological surface states in metallic Bi$_2$Te$_3$ topological insulator}
\author{K. Shrestha$^{1,}$}
\email[Corresponding E-mail:]{keshav.shrestha@inl.gov}
\author{M. Chou$^{2}$, D. Graf$^3$, H. D. Yang$^{4}$, B. Lorenz$^{5}$ and C. W. Chu$^{5,}$$^{6}$}
\affiliation{$^{1}$Idaho National Laboratory, 2525 Fremont Ave, Idaho Falls, Idaho, 83402, USA}
\affiliation{$^{2}$Department of Materials and Optoelectronic Science, National Sun Yat Sen University, Taiwan}
\affiliation{$^{3}$National High Magnetic Field Laboratory, Florida State University, Tallahassee, Florida 32306, USA}
\affiliation{$^{4}$Department of Physics, National Sun Yatsen University, Taiwan}
\affiliation{$^{5}$TCSUH and Department of Physics, University of Houston, 3201 Cullen Boulevard, Houston, Texas 77204, USA}
\affiliation{$^{6}$Lawrence Berkeley National Laboratory, 1 Cyclotron Road, Berkeley, California 94720, USA}

\begin{abstract}
    Weak antilocalization (WAL) effects in Bi$_2$Te$_3$ single crystals have been investigated at high and low bulk charge carrier concentrations.
     At low charge carrier density the WAL curves scale with the normal component of the magnetic field, demonstrating the dominance of topological surface states in magnetoconductivity. At high charge carrier density the WAL curves scale with neither the applied field nor its normal component, implying a mixture of bulk and surface conduction. WAL due to topological surface states shows no dependence on the nature (electrons or holes) of the bulk charge carriers. The observations of an extremely large, non-saturating magnetoresistance, and ultrahigh mobility in the samples with lower carrier density further support the presence of surface states. The physical parameters characterizing the WAL effects are calculated using the Hikami-Larkin-Nagaoka formula. At high charge carrier concentrations, there is a greater number of conduction channels and a decrease in the phase coherence length compared to low charge carrier concentrations. The extremely large magnetoresistance and high mobility of topological insulators have great technological value and can be exploited in magneto-electric sensors and memory devices.

\end{abstract}


\maketitle
\section{Introduction}
The quantum interference of electrons in solids experiencing a backscattering event, which limits the electrical conductivity, has been of interest for many years. The resistance in systems with scattering centers increases with decreasing temperature due to increasing phase coherence in backscattering processes where the electron moves through a loop of scattering sites clockwise, as well as counter-clockwise, and the constructive superposition of the scattering amplitudes results in an enhancement of the resistivity. The phase coherence in this additive scattering process, coined weak localization (WL), is destroyed in an external magnetic field, resulting in a characteristic negative magnetoresistance\cite{Chakravarty1986}. On the other hand, in systems with strong spin-orbit interaction, the spin of the electron is tied to the momentum. This results in a rotation of the spins along the backscattering path and the phase difference between clockwise and counter-clockwise trajectories amounts to 360$^o$. Rotating a spin by 360$^o$ changes the sign of the scattering amplitude and the two scattering events experience a destructive interference, reducing the resistance in materials with strong spin-orbit coupling [weak antilocalization (WAL)]. The destructive interference effects are diminished in magnetic fields resulting in a positive magnetoresistance, as observed in many metallic compounds with large spin-orbit interaction\cite{Bergmann1982}.\\
\indent WL and WAL are more pronounced in two-dimensional (2D) systems since the probability of two interfering scattering pathways, distinguished only by time reversal, is larger in lower dimensions. Therefore, surface states of topological insulators (TIs) have been studied recently with respect to the WAL effects. The surface conduction can be easily studied when the bulk of the TI is insulating and does not contribute to the conductivity. In many TIs the study of transport from topological surface states is impeded by a significant bulk contribution to the overall conduction due to defects, impurities, or intersite occupancy. This makes it difficult to extract the topological properties of the surface states by employing magnetotransport methods. However, in a recent investigation of metallic, hole-doped Bi$_2$Se$_{2.1}$Te$_{0.9}$ and Sb$_2$Te$_2$Se, we have shown that quantum oscillations from topological surface states can be resolved well despite the interference with bulk conduction\cite{Shrestha2014}$^,$\cite{KShrestha}$^,$\cite{KShrestha1}. This motivated us to search for other physical phenomena related to topological surface states in TIs with bulk metallic properties.\\
\indent Bismuth telluride Bi$_2$Te$_3$ is shown to be a three dimensional topological insulator by both theoretical and experimental studies\cite{Chen2013}$^{,}$\cite{He2011}\cite{Zhang2009}. Recently Qu $et$ $al.$\cite{Qu2010} observed that every single crystal of Bi$_2$Te$_3$ can have different physical properties such as bulk carrier concentrations, carrier types, etc. even if they are grown in the same batch. They reported that this might be due to a weak or critical composition gradient during the crystal growth process. In this work, we report the WAL effect in Bi$_2$Te$_3$ single crystals with metallic bulk conductivity and different carrier densities with hole as well as electron nature. Three single crystals, S1, S2, and S3, are selected from the same boule of crystals. People have used either angle dependence of quantum oscillations or WAL curves to identify the presence of topological surface states\cite{Qu2010}$^{,}$\cite{Bao2012}. In our study, even though samples S1 and S2 show quantum oscillations under fields above 7 T, S3 did not show any signature of quantum oscillations in the field up to 31 T. However, all of those samples show WAL effect at low temperature. That is why we have used angle dependence of WAL curves to detect the presence of topological surface states in all samples. For two crystals with lower bulk carrier density (S1, electron-, and S3, hole-doped) we find the clear signature of WAL from 2D surface states. The dependence of the magnetoresistance on the magnetic field angle with the sample's surface clearly proves the 2D character of the WAL effect. Another single crystal, S2, with higher bulk carrier density does not show a decisive dependence on the field angle, which indicates that WAL effects from surface and bulk states interfere with one another and cannot be unequivocally separated.
\section{Experimental}
\indent Single crystals of Bi$_2$Te$_3$ were grown by a home-made resistance-heated floating zone furnace (RHFZ). The starting materials of Bi$_2$Te$_3$ were prepared in the stoichiometric mixture of high purity elements, Bi (99.995\%) and Te (99.995\%). The mixture was sintered at 820$^{o}$C for 20 hours and then slowly cooled down to room temperature in an evacuated quartz tube. This material was then used as a feeding rod for the RHFZ experiment. The as-grown crystals were cleaved naturally with a silvery shining mirror-like surface.  RHFZ is a method similar to directional solidification, in which a small region of the feeding rod gets melted, and this molten zone is moved along the feeding rod. The molten region will move impurities to one end of the feeding rod and as it moves through the ingot, and this leaves a wake of purer material solidified behind it. The impurities concentrated in the melt have an appreciable concentration difference between the solid and liquid phases at equilibrium. It was found that a topological insulator crystal grown using the RHFZ method has a better crystalline uniformity than one grown by the traditional vertical Bridgman method. All three single crystals (S1, S2, and S3) studied here are selected from the same batch.\\
\indent Magnetoresistance measurements in magnetic fields up to B=7 T were carried out using the ac-transport option of the physical property measurement system (PPMS, Quantum Design). A very thin, rectangular piece of Bi$_2$Te$_3$ was peeled using Scotch tape and attached to a magnesium oxide (MgO) substrate using GE-varnish. Typical dimensions for these crystals are $\approx$ 3mm$\times$2mm$\times$0.05mm. The sample was covered with a plastic mask and six gold contacts were sputtered on the freshly cleaved sample surface. The sample was then mounted on a rotation platform for measuring longitudinal and Hall resistances.  High field MR measurements at the National High Magnetic Field Laboratory (NHMFL) were performed using conventional lock-in techniques. A sample was mounted onto the rotating platform of a probe designed at NHMFL. The probe was inserted into the sample space of a $^3$He Oxford cryostat, which is installed in a bore of a resistive magnet with a maximum field of 35 T.  A Keithley (6221) current source excites the sample at a fixed frequency and a lock-in amplifier (SR-830) was used to measure longitudinal and Hall voltages at the same frequency. The sample position with respect to applied field was calibrated by using a Hall sensor.
\section{Results and Discussion}
\begin{figure}
  \centering
  \includegraphics[width=1.0\linewidth]{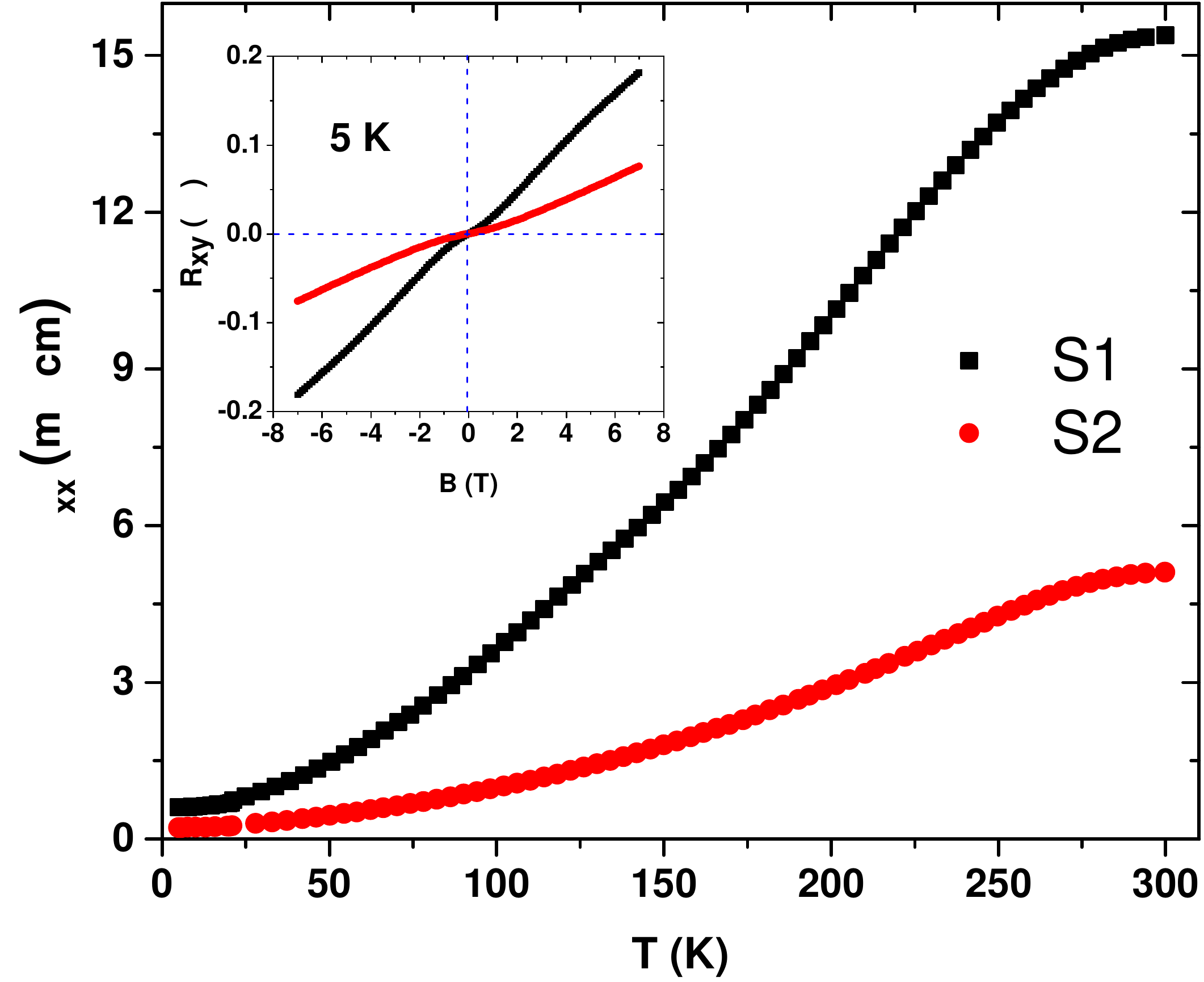}
  \caption{Temperature dependence of longitudinal resistivity for Bi$_2$Te$_3$ single crystals, S1 (black squares) and S2 (red circles). The inset shows the Hall resistance, $R_{xy}$, versus $B$ at T = 5 K.}\label{Resistivity}
\end{figure}
\indent Figure [1] shows the temperature dependence of longitudinal resistivity, $\rho_{xx}$, for S1 and S2. Both samples show a metallic behavior below room temperature. The resistivity of S1 is higher than that of S2 throughout the entire temperature range. The residual resistivity ratios, defined as RRR=$\rho_{xx}$(300 K)/$\rho_{xx}$(5 K), where $\rho_{xx}$(300 K) and $\rho_{xx}$(5 K) are the resistivity values at T=300 and 5 K, respectively, of S1 and S2 are calculated to be 25 and 23, respectively. Such large RRR values reflect the high crystalline quality of S1 and S2. The inset to figure [1] displays the Hall resistance, $R_{xy}$, versus $B$ of S1 and S2 at T=5 K. Non-linear field dependence of $R_{xy}$ at $B$=0 suggests the existence of a multiband effect (electron and hole bands), as observed in other bismuth-based topological systems \cite{Shrestha2014}$^{,} $\cite{Kushwaha2016}. The positive slope of the $R_{xy}(B)$ curve implies the dominance of $p$-type bulk charge carriers in S1 and S2. For the sake of simplicity, we have used the single-carrier Drude band model ($N$=$B/\left|e\right|\rho_{xy}$, where $\rho_{xy}$ is the Hall resistivity and $\left|e\right|$ is electron's charge) for the calculation of the bulk charge carrier density. At T=5 K, we have estimated bulk charge carriers of 6$\times$10$^{17}$ and 3$\times$10$^{18}$ cm$^{-3}$ for S1 and S2, respectively, giving sample S2 almost 5 times as many bulk charge carriers as S1.\\
\begin{figure}
  \centering
  \includegraphics[width=1.0\linewidth]{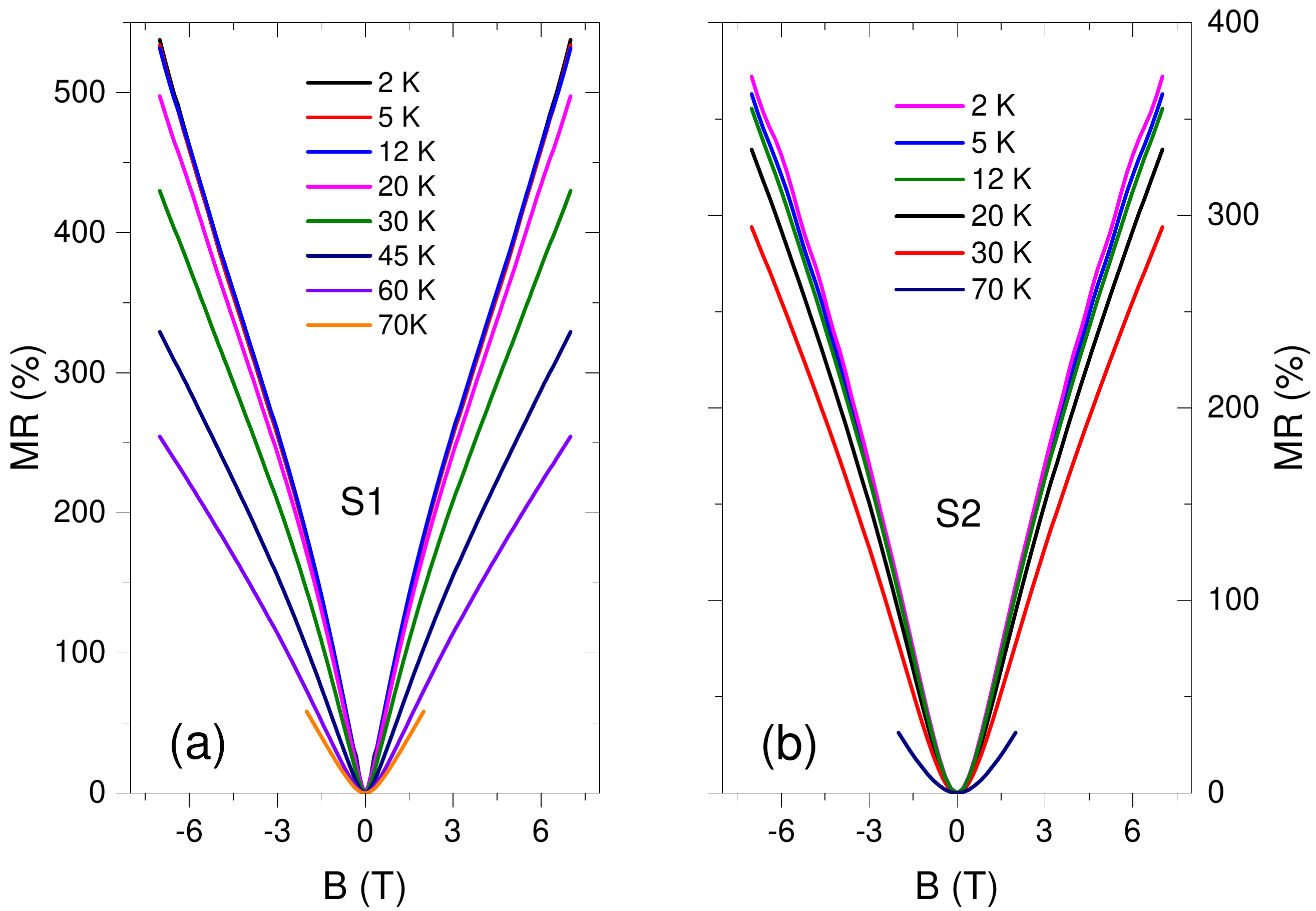}
  \includegraphics[width=1.0\linewidth]{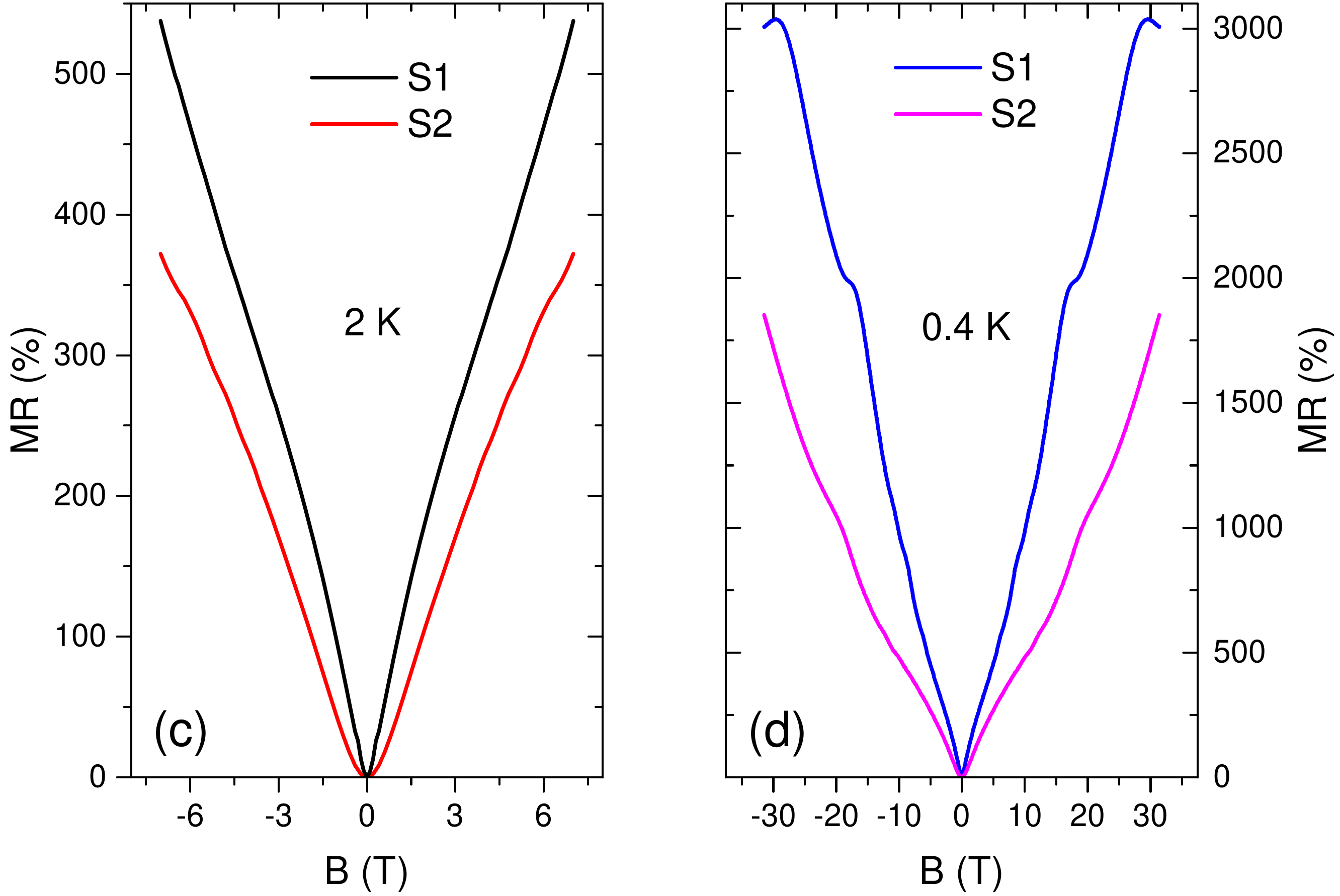}
  \caption{Temperature dependence of MR up to 7 T for (a) S1, and (b) S2. (c) Comparison of the MR curves for S1 and S2 at 2 K. (d) The MR curves of S1 and S2 measured in high fields up to 34 T at T=0.4 K.}\label{WALA}
\end{figure}
\indent For the magnetoresistance (MR) measurements of S1 and S2, we have calculated MR as a percentage, defined as MR=\big[$R_{xx}(B)$/$R_{xx}(0)$-1\big]$\times$100\%, where $R_{xx}(0)$ and $R_{xx}(B)$ are resistances at zero and $B$ applied field, respectively. Figure [2(a)] shows MR of S1 with $B$ perpendicular to the sample surface (a-b plane). At $T$=2 K, the MR curve of S1 increases linearly with $B$ and reaches 540\% at 7 T. At given field $B$=7 T, this MR value is significantly higher than the previous reports of MR=240\% by Wang $et$ $al.$,\cite{Wang2012} and 90\% by Qu $et$ $al.$\cite{Qu2010}. With increasing temperature, MR remains unchanged up to 12 K, and then decreases rapidly with a further increase in temperature. At $T$=60 K, MR reaches 250\% at 7 T, which is almost $\frac{1}{2}$ of the value at $T$=2 K. The MR curve shows the parabolic field dependence at higher temperature and low magnetic fields (see the MR curve at $T$=70 K in figure [2(a)]). MR of S2 is displayed in figure [2(b)]. MR increases to 370\% at $T$=2 K under 7 T. This value is relatively lower (almost 1.5 times smaller) than that of S1. The MR curves of S2 display temperature dependence similar to that of S1. Furthermore, MR of S1 and S2 shows a sharp cusp-like feature at low temperature, which indicates the existence of the WAL effect \cite{He2011}$^{,}$ \cite{Chen2014}. The cusp-like feature of S1 is sharper than that of S2, as shown in figure [2(c)]. This could be due to the contribution of more bulk states\cite{Bao2012} to the MR of S2, which we will later discuss in detail.\\
\indent MR of S1 and S2 increases with magnetic field and does not show any sign of saturation. In order to investigate further, we have carried out MR measurements in dc magnetic fields up to 34 T at the National High Magnetic Field Laboratory (NHMFL), in Tallahassee, Florida. S1 shows a massive increase in MR, i.e. 3300\% at 0.4 K under 34 T, and still displays no signature of saturation, as shown in figure [2(d)]. This is the first time such a large MR value is observed in bismuth-based topological systems, and it is comparable with that of Dirac semimetal Cd$_3$As$_2$\cite{Narayanan2015}$^{,}$\cite{Liang2015}. However, MR of S2 reaches 1850\% under 31 T at 0.4 K, which is almost half that of S1. Large MR is usually linked with high mobility, as observed in many Weyl and Dirac materials \cite{Shekhar2015}$^{,}$ \cite{Liang2015}. We have used the simple Drude model [$\mu(T)$=$R_{H}(T)/\rho_{xx}(T)$, where $R_{H}(T)$ is the Hall coefficient at temperature T] for estimating the effective mobility of S1 and S2. From our calculations, we have found $\mu$=4.5$\times$10$^{4}$ and 3.6$\times$10$^{3}$ cm$^2$V$^{-1}$S$^{-1}$ for S1 and S2 at 5 K, respectively. The high mobility of S1 is comparable to that of Bi$_2$Te$_3$ samples that show topological surface states\cite{Qu2010}, and even to that of a Cd$_3$As$_2$ sample\cite{He2014}. The mobility of S2 is one order of magnitude lower than that of S1. The linear non-saturating MR was proposed by Abrikosov\cite{Abrikosov1998} in systems that show a linear dispersion relation. Due to the linear dispersion of surface states, a large non-saturating MR is expected to be seen. Also, due to symmetry protection, the surface electrons are robust against impurity and have ultrahigh mobility. Thus, observation of these properties in S1 suggests that the WAL effect could be a result of topological surface states. In order to clarify the origin of WAL in S1 and S2, we have performed the magnetoresistance measurements at different tilt angles.\\
\indent Figure [3(a)] shows magnetoconductivity (MC), defined as $\sigma(B,\theta)$=$R_{xx}(0)$/$R_{xx}(B,\theta)$, measured along different tilt angles, $\theta$, at T=0.4 K. Here, $\theta$ is defined as the angle between magnetic field and current directions (see the inset to figure 3 (a)). All of the MC curves of S1 merge together when they are plotted as a function of the normal component of field, $Bsin\theta$, as shown in figure [3(b)]. This provides strong evidence that topological surface states dominate over the bulk states \cite{Lee2012}$^{,} $\cite{Chen2011} in the MC of S1.\\
\begin{figure}
  \centering
  \includegraphics[width=1.0\linewidth]{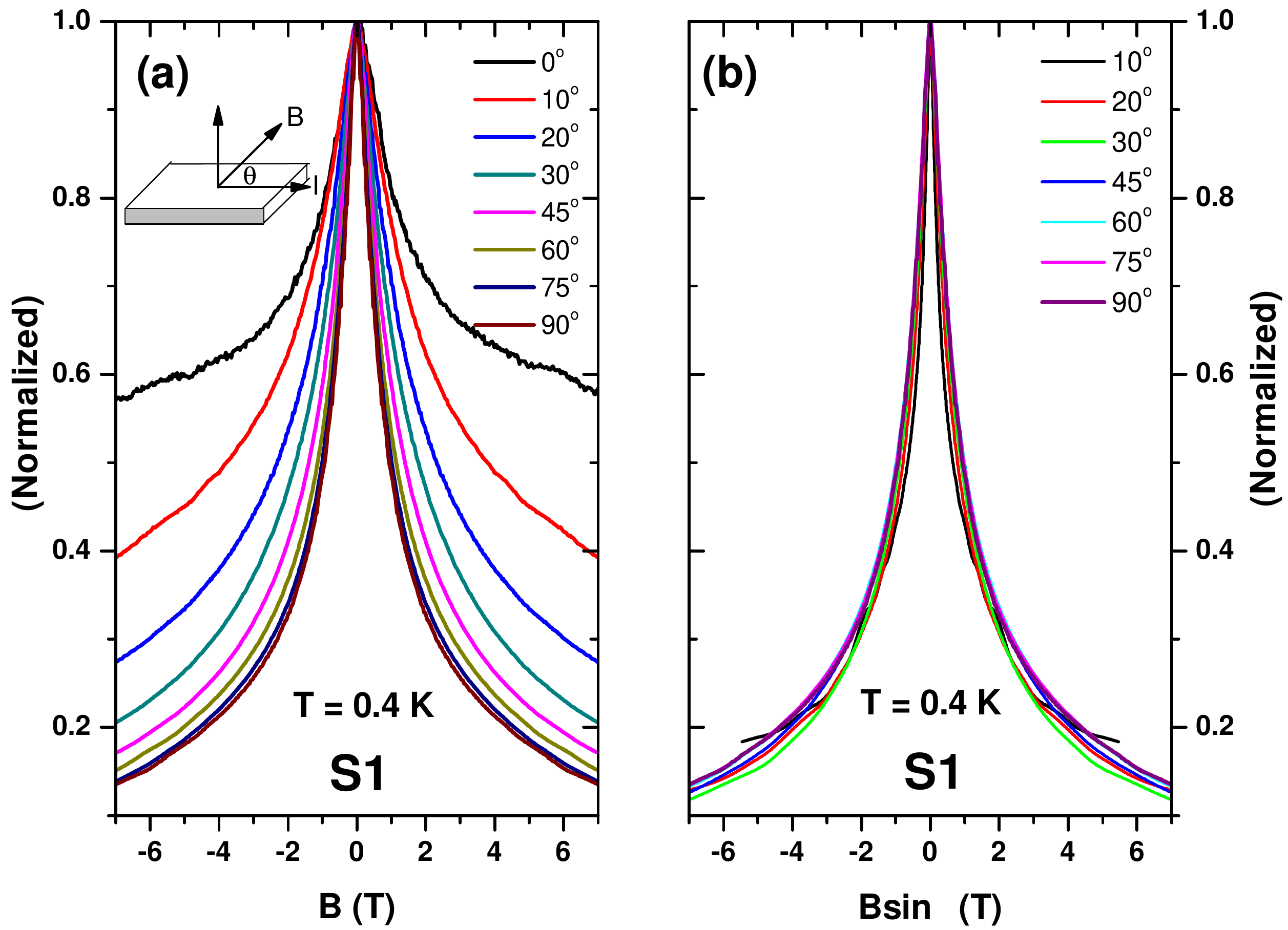}
  \includegraphics[width=1.0\linewidth]{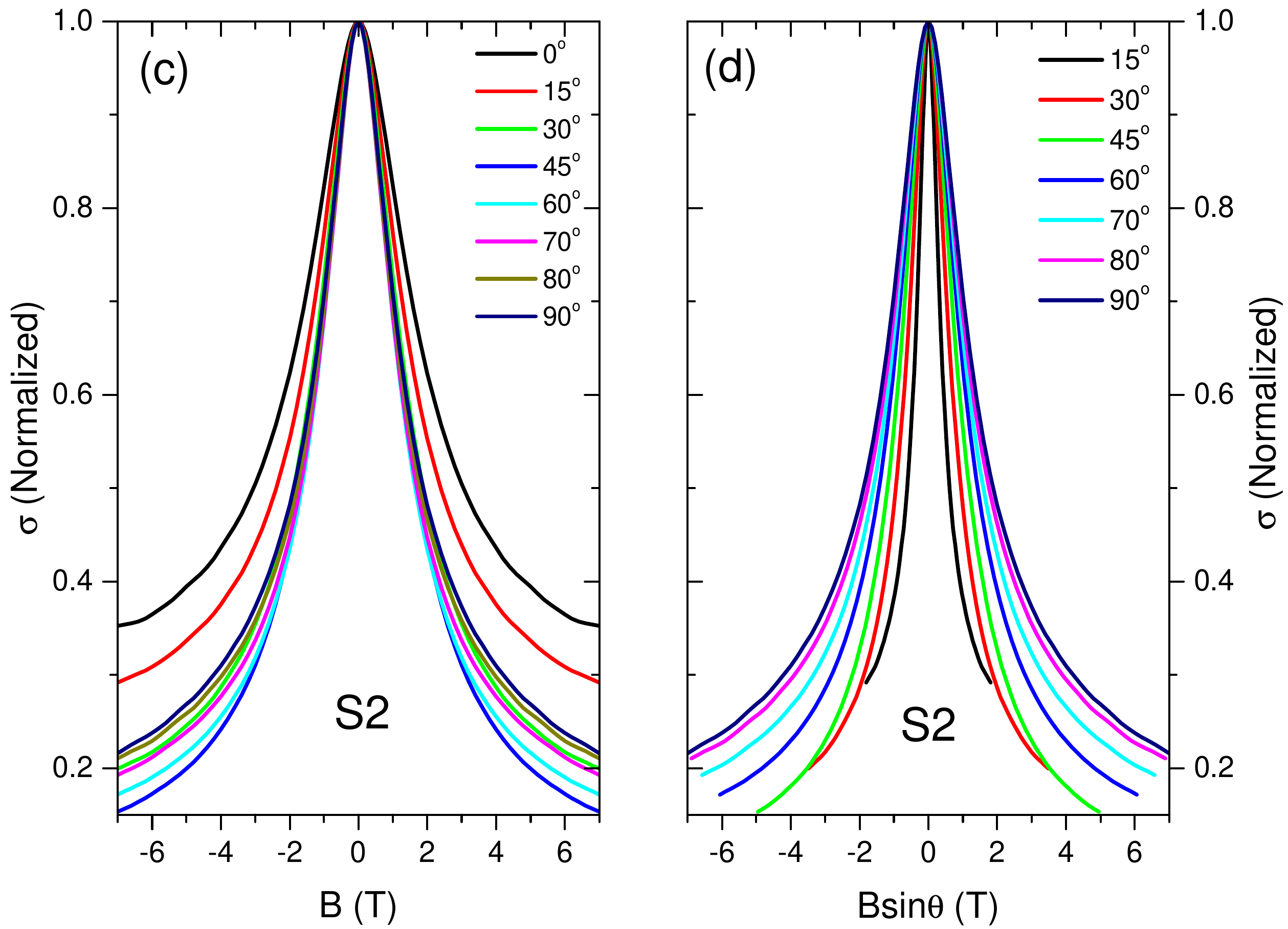}
  \caption{MC curves of S1 as a function of (a) $B$ and (b) $B$sin$\theta$, under fields up to 7 T at T = 0.4 K. Inset in figure (a): Sample configuration for MC measurements. MC curves of S2 as a function of (c) $B$ and (d) $B$sin$\theta$ under fields up to 7 T at T = 5 K.}\label{WALA}
\end{figure}
\indent We have also carried out similar angle-dependent MC experiments on S2 under fields up to 7 T. Figure [3(c)] shows the MC curves of S2 in the field range (-7 to 7 T) at T = 5 K. Initially, MC shows a strong angle dependence while increasing the tilt angle from $\theta$=0$^o$ to 30$^o$; however, it weakly changes with further increase in $\theta$. Also, the MC curves disperse when plotted as a function of the normal component $Bsin\theta$, as shown in figure [3(d)]. If WAL is caused mainly by the spin-orbit coupling in a 3D bulk channel, MC is independent of $\theta$. This scenario of the MC curves, which scale neither with $B$ nor normal component $Bsin\theta$, strongly suggests the presence and superposition of two contributions to MC, one from surface and one from bulk states, which is reasonable since S2 has a higher bulk carrier density than S1. \\
\indent From the above discussion, we have confirmed that the WAL effect in S1 is due to topological surface states. Here, both S1 and S2 show metallic behavior, and have $p$-type bulk charge carriers. The only difference is that S1 has fewer bulk charge carriers than S2. This indicates that the carrier density is an important factor for the observation of topological surface states in MC measurements. However, whether the domination of topological surface states over the bulk states in MC depends on the nature of the bulk carriers remains a question. To answer this, we have selected a third single crystal, S3, with electron-like carriers for a comparative study. \\
\begin{figure}
  \centering
  \includegraphics[width=1.0\linewidth]{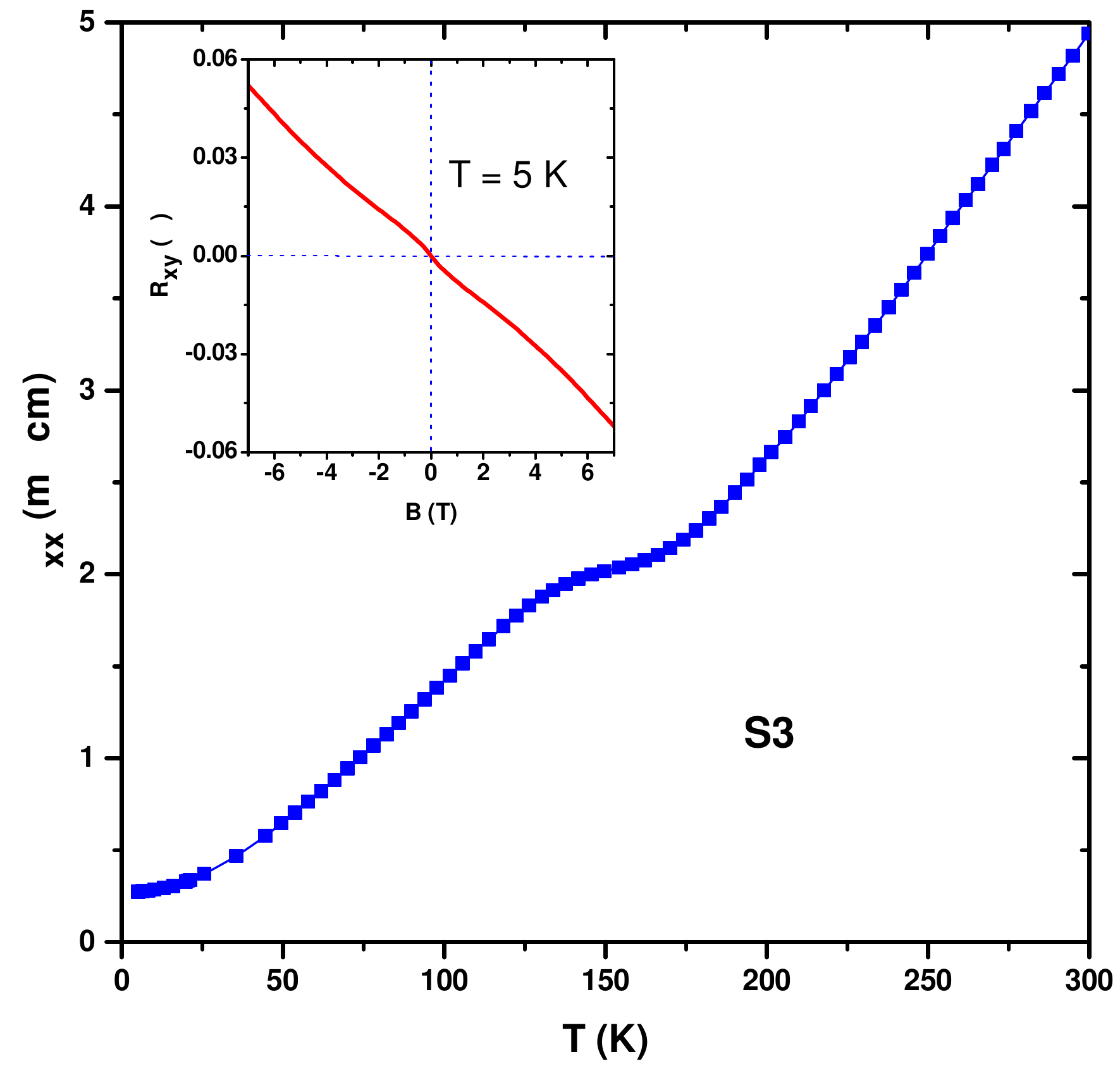}
  \caption{Temperature dependence of longitudinal resistivity for Bi$_2$Te$_3$ single crystal, S3. The inset shows the Hall resistance, $R_{xy}$, versus $B$ at T=5 K.}\label{Fig5}
\end{figure}
\indent The resistivity of S3 also shows metallic behavior below room temperature, as shown in figure 4. At T=150 K, there is a slight upturn in the resistivity, but it decreases with further cooling. Qu $et$ $al.$\cite{Qu2010} have also observed a resistivity increase starting at T=150 K in Bi$_2$Te$_3$ single crystals. The RRR value, 18, of S3 is comparable to those of S1 and S2. The negative slope of $R_{xy}$ versus $B$ (see inset to figure [SP2] in the Supplemental Materials) confirms the presence of $n$-type bulk charge carriers. From the Hall data analyses, the bulk carrier density is estimated to be 1.0$\times$10$^{18}$ cm$^{-3}$. This value lies in between the carrier densities of S1 and S2. \\
\begin{figure}
  \centering
  \includegraphics[width=0.95\linewidth]{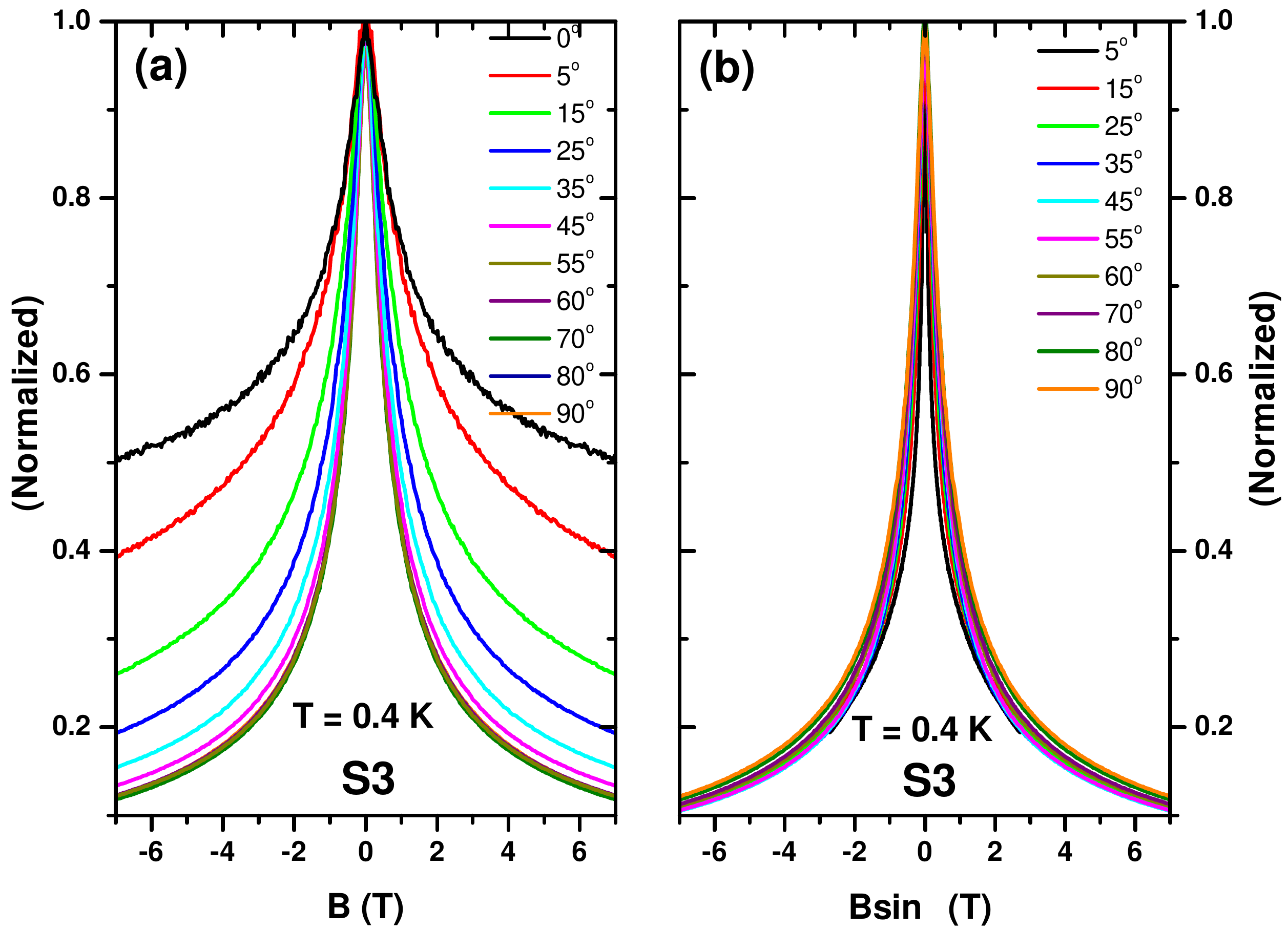}
  \includegraphics[width=0.95\linewidth]{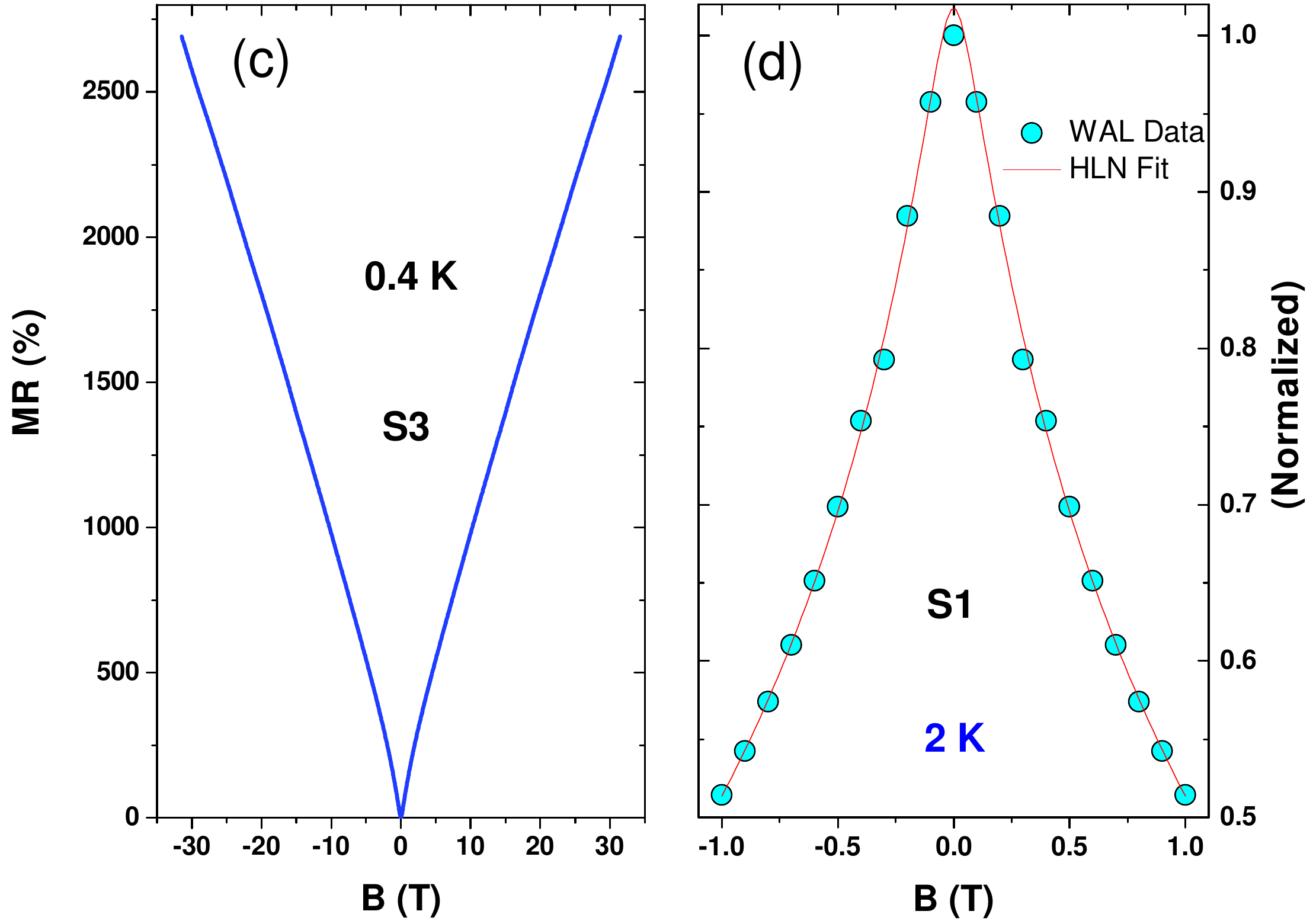}
  \includegraphics[width=0.95\linewidth]{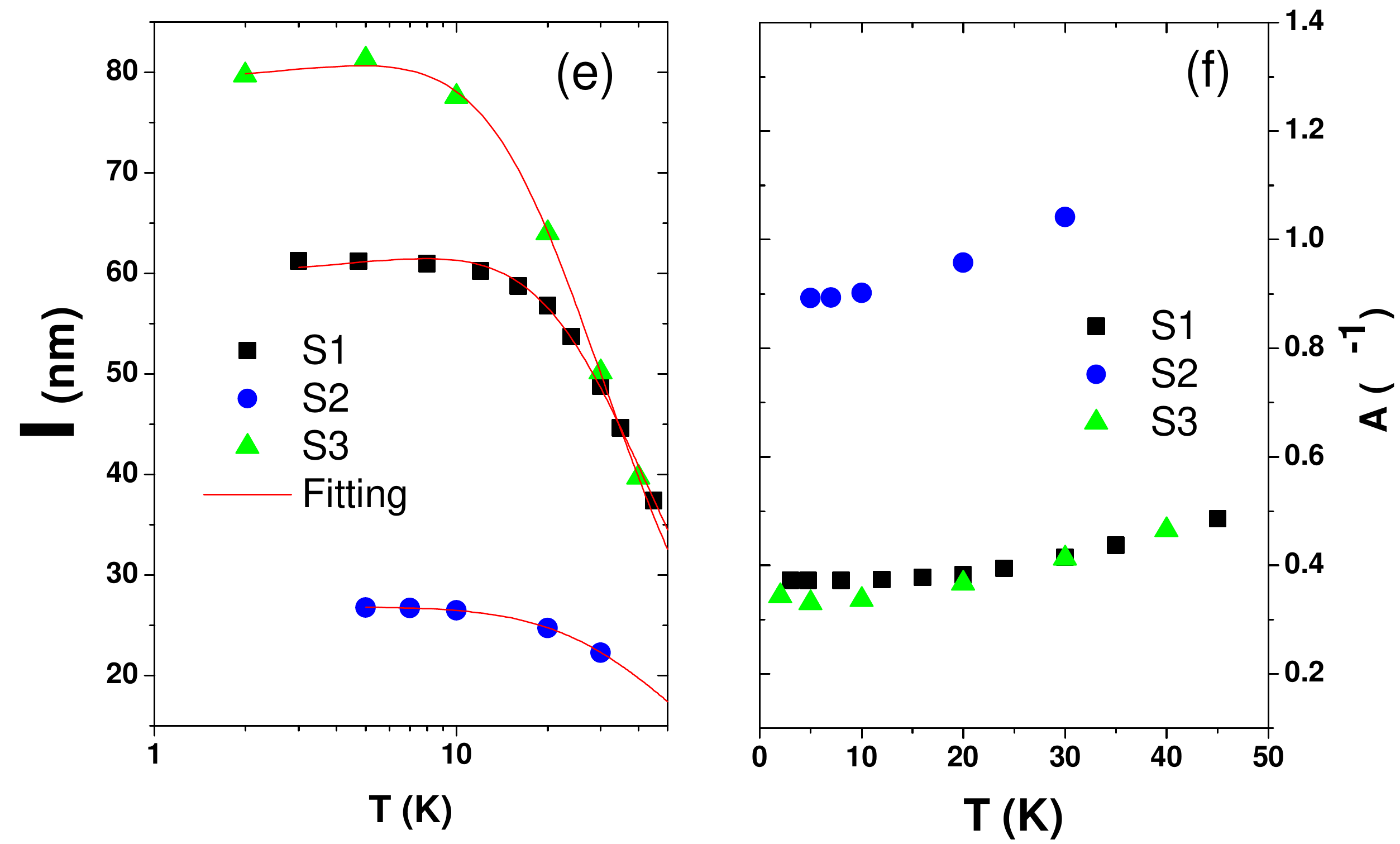}
  \caption{MC curves of S3 measured in fields up to 7 T at 0.4 K, plotted as a function of (a) $B$, and (b) $B$sin$\theta$. (c) MR curve of S3 under fields up to 31 T at T = 0.4 K. (d) The HLN formula fitting (red curve) of S1 within (-1 to 1 T) range. (e) Temperature dependence of $l_{\phi}$ for S1, S2, and S3. Solid red curves are fittings using equation (\ref{Lphi}). (f) $A$ values of S1, S2, and S3 at different temperatures.}\label{HLN}
  \end{figure}
\indent Figure [5(a)] shows the MC curves of S3 along different $\theta$ at T=0.4 K. A cusp-like feature at low magnetic field shows the existence of the WAL effect in S3 as well. The WAL curves collapse together with the normal component $Bsin\theta$ (figure [5(b)]), confirming the dominance of topological surface states in the MC of S3. Similar to the S1 and S2 samples, we have also determined MR and mobility of S3. MR of S3 increases to 2700\% under 31 T at T=0.4 K, and does not show any indication of saturation, as shown in figure [5(c)]. Using the Hall coefficient and resistivity data, we have estimated mobility $\mu$=1.3$\times$10$^{4}$ cm$^2$V$^{-1}$S$^{-1}$. These observations in sample S3 confirm that the domination of topological surface states in MC depends only on the carrier density and is independent of the nature of the charge carriers. \\
\indent We have also estimated several physical parameters that characterize the WAL effect by employing the Hikami-Larkin-Nagaoka (HLN) formula \cite{Hikami1980}. The HLN formula has already been used to describe the WAL effect in topological systems, for example topological thin films\cite{Taskin2012} and single crystals\cite{Shekhar2014}. According to the HLN formula, MC can be expressed as,\\
\begin{equation}\label{Hikami}
\sigma(B)=-A\bigg[\Psi\bigg(\frac{1}{2}+\frac{\hbar}{4el^{2}_{\phi}B}\bigg)-ln\bigg(\frac{\hbar}{4el^{2}_{\phi}}\bigg)\bigg].
\end{equation}\\
\noindent Here $\Psi$ is the digamma function, and $l_{\phi}$ is the phase coherence length, which is the distance traveled by an electron before its phase is changed. The parameter $A$=$\alpha\frac{e^2}{2\pi^{2}\hbar}$ with $\alpha$=1/2 per conduction channel. Thus, $A$ represents the number of conduction channels present in a sample. Using equation (1) with our experimental data, the fitting parameters $l_{\phi}$ and $A$ can be determined.\\
\indent Figure [5(d)] shows the HLN fitting to the MC curve of S1 in low field range (-1 to 1 T). Similar HLN fitting was performed for the MC data of S2 and S3 to determine $A$ and $l_{\phi}$. The $l_{\phi}$ values of S1, S2, and S3 at different temperatures are presented in figure [4(e)]. Temperature dependence of $l_{\phi}$ can  be described as\cite{Zhu2012}$^{,} $\cite{Xu2014a}
\begin{equation}\label{Lphi}
\frac{1}{l_{\phi}^{2}(T)}=\frac{1}{l_{\phi}^{2}(0)}+A_{ee}T^{p_{1}}+A_{ep}T^{p_{2}},
\end{equation}
 where, $l_{\phi}$(0) is the zero-temperature phase coherence length, and $A_{ee}T^{p_{1}}$ and $A_{ep}T^{p_{2}}$ represent the contributions from the electron-electron and electron-phonon interactions, respectively. Equation (2) is fitted to the temperature dependence of $l_{\phi}$ with $p_{1}$=1 and $p_{2}$=2, shown by solid curves in figure [4(e)]. The fitting parameters $l_{\phi}$(0), $A_{ee}$, and $A_{ep}$ of S1, S2, and S3 are presented in table [1]. The $l_{\phi}$(0) values of S1 and S3 are comparable, but are almost 3 times as large as that of S2. The larger $l_{\phi}$(0) values in S1 and S3 are comparable to those of other topological systems\cite{Roy2013}$^{,}$\cite{Liu2011}.\\
\indent Figure [5(f)] shows the parameter $A$ as a function of temperature. The $A$ values of S1, S2, and S3 are on the order of 10$^0$, which is nearly 10$^5$ times larger than that for two dimensional systems. These observations are also seen in other topological single crystals\cite{Xu2014a}$^{,}$ \cite{Checkelsky2009} and this discrepancy could be due to the contribution from the dominant bulk channels. The $A$ values remain nearly constant with temperature up to 45 K, suggesting the presence of a fixed number of conduction channels in S1, S2, and S3. Moreover, the value of $A$ for S1 is comparable to that for S3, but is almost half of the value for S2. Since the value of $A$ is a measure of the number of conduction channels present in a sample, S1 and S3 have a smaller number of conduction channels as compared to S2.

\begin{table}[h!]
  \centering
  \begin{tabular}{|c|c|c|c|}
  \hline
   Samples & $l_{\phi}$(0) (nm) & $A_{ee}$ (nm$^{-2}$) & $A_{ep}$ (nm$^{-2}$)\\
   \hline
   S1 & 59 & -5.13$\times$ 10$^{-6}$ & -3.25$\times$ 10$^{-7}$ \\
    \hline
    S2& 26 & -6.17$\times$ 10$^{-6}$ & 8.83$\times$ 10$^{-7}$ \\
    \hline
    S3& 78 & 3.75$\times$ 10$^{-6}$ & 3.88$\times$ 10$^{-7}$ \\
    \hline
  \end{tabular}
  \caption{Comparison of the parameters $l_{\phi}$(0), $A_{ee}$, and $A_{ep}$ for S1, S2, and S3 samples. The larger coherence lengths in S1 and S3 support the observation of surface states in MC of those samples.}\label{t1}
\end{table}

\section{Summary}
\indent In summary, we have studied the magnetoresistance of three Bi$_2$Te$_3$ single crystals, all having bulk metallic properties but with different concentrations and nature of charge carriers. Whereas all of the samples show very large MR and high mobility, the two crystals with lower carrier density (one electron and one hole like) exhibit the largest MR and mobility, comparable with values observed in the Dirac semimetal Cd$_3$As$_2$. We have also demonstrated that large MR and high mobility in Bi$_2$Te$_3$ depend only on the bulk carrier concentration and are independent of the nature of the charge carriers. The cusp of MR for the samples with low carrier density reflects the characteristics of the WAL effects. The dependence of MC on the angle of the magnetic field with the sample's surface demonstrates that MR is dominated by topological surface conduction in the low carrier density samples. The third sample with higher carrier density shows an interference of surface and bulk effects on MR, as can be expected with increasing bulk carrier number. The Hikami-Larkin-Nagaoka formula is used to calculate different physical quantities that characterize the observed WAL effects. The larger phase coherence length and higher electrical mobility of, and the smaller number of conduction channels present in, the crystals with lower carrier density, confirm that topological surface states dominate the magnetic field effect on the overall conduction in these samples.

\section*{acknowledgements}
This work is supported in part by the U.S. Air Force Office of Scientific Research, the T. L. L. Temple Foundation, the J. J. and R. Moores Endowment, and the State of Texas through the TCSUH. The work at NHMFL is supported by the NSF co-operative Agreement No. DMR-1157490 and the State of Florida. K.S. acknowledges the Department of Energy, Office of Basic Energy Sciences, Materials Sciences, and Engineering Division and through grant DOE FG02-01ER45872.

\bibliography{WAL}

\end{document}